\documentclass[11pt,twoside]{article}

\usepackage{asp2014}

\aspSuppressVolSlug
\resetcounters

\begin{document}

\title{Requirements analysis for HPC\&HTC infrastructures integration in ESCAPE Science Analysis Platform}

\author{Sara~Bertocco$^1$, David~Goz$^1$, Stefano~Alberto~Russo$^1$, Marco~Molinaro$^1$, and Giuliano~Taffoni$^1$}
\affil{$^1$INFN - Istituto Nazionale di Fisica Nucleare, Trieste, Italy; \email{sara.bertocco@inaf.it}}

\paperauthor{Sara~Bertocco}{sara.bertocco@inaf.it}{orcid.org/0000-0003-2386-623X}{INAF}{Osservatorio Astronomico di Trieste}{Trieste}{}{34143}{Italy}
\paperauthor{David Goz}{david.goz@inaf.it}{orcid.org/0000-0001-9808-2283}{INAF}{Osservatorio Astronomico di Trieste}{Trieste}{Trieste}{34143}{Italy}
\paperauthor{Stefano~Alberto~Russo}{setefano.russo@inaf.it}{orcid.org/0000-0003-4487-6752}{INAF}{Osservatorio Astronomico di Trieste}{Trieste}{}{34143}{Italy}
\paperauthor{Marco~Molinaro}{marco.molinaro@inaf.it}{orcid.org/0000-0001-5028-6041}{INAF}{OATs}{Trieste}{}{34143}{Italy}
\paperauthor{Giuliano~Taffoni}{giuliano.taffoni@inaf.it}{orcid.org/0000-0002-4211-6816}{Istituto Nazionale di Astrofisica}{Osservatorio Astronomico di Trieste}{Trieste}{}{34143}{Italy}



\begin{abstract}
ESCAPE (European Science Cluster of Astronomy and Particle physics ESFRI research infrastructures) is a project to set up a cluster of ESFRI (European Strategy Forum on Research Infrastructures) facilities for astronomy, astroparticle and particle physics to face the challenges emerging through the modern multi-disciplinary data driven science. 
One of the main goal of ESCAPE is the building of ESAP (ESFRI Science Analysis Platform), a science platform for the analysis of open access data available through the EOSC (European Open Science Cloud) environment. ESAP will allow EOSC researchers to identify and stage existing data collections for analysis, share data, share and run scientific workflows.
For many of the concerned ESFRIs and RIs, the data scales involved require significant computational resources (storage and compute) to support processing and analysis. The EOSC-ESFRI science platform therefore must implement appropriate interfaces to an underlying HPC (High Performance Computing) or HTC (High Throughput Computing) infrastructure to take advantage of it. 
This poster describes the analysis done to identify the main requirements for the implementation of the interfaces enabling the ESAP data access and computation resources integration in HPC and HTC computation infrastructures in terms of authentication and authorization policies, data management, workflow deployment and run.
\end{abstract}
%
%
%
%
\section{The ESCAPE Project}
ESCAPE (European Science Cluster of Astronomy and Particle physics ESFRI research
infrastructures) is a project funded in the European Community research program H2020 to set up a cluster of ESFRI (European Strategy Forum on Research Infrastructures) facilities for
astronomy, astroparticle and particle physics to face the challenges emerging through the modern multi-disciplinary data driven science. This cluster aims to state a functional
connection between the interested ESFRI projects and the EOSC (European Open Science Cloud)
providing tools and solutions according to FAIR (Findable, Accessible, Interoperable and
Reusable) principles. 
\section{The ESAP: ESFRI Science Analysis Platform}

One of the main goal of ESCAPE is the building of ESAP (ESFRI Science Analysis Platform), a flexible and expandable science platform for the analysis of open access data available through the EOSC environment. ESAP will provide users with the capability to identify, access and combine data from multiple, large and distributed collections and to stage them for subsequent processing and analysis. ESAP aims moreover to give its users the possibility to use a wide-range of software tools and packages developed by and in support of the ESFRIs, bringing their own custom workflows to the platform, and taking advantage of the underlying high performance or high throughput computing infrastructure to execute those workflows. \\
{\bf General use case scenario:} \\
--- the user connects to ESAP \\
--- the user exploits ESAP VO (Virtual Observatory) tools powered functionalities to search data in large and distributed collections \\
--- the user exploits ESAP Open-Source Scientific Software and Service Repository to select software and/or services to analyse/perform tasks on selected data \\
--- the user selects a computing center, HPC or HTC, among the available integrated in ESAP \\
--- the data are staged in the selected data center \\
--- the user performs his analysis work, interactively or in batch mode, depending from the services/tools selected. \\
There are two different scenarios for what concerns the data staging: in one the data are in the same data center of the computation resources and in another the data are on a different location than the computation resources (see Figures \ref{use_case_fig1}-\ref{use_case_fig2}).  
\articlefigure{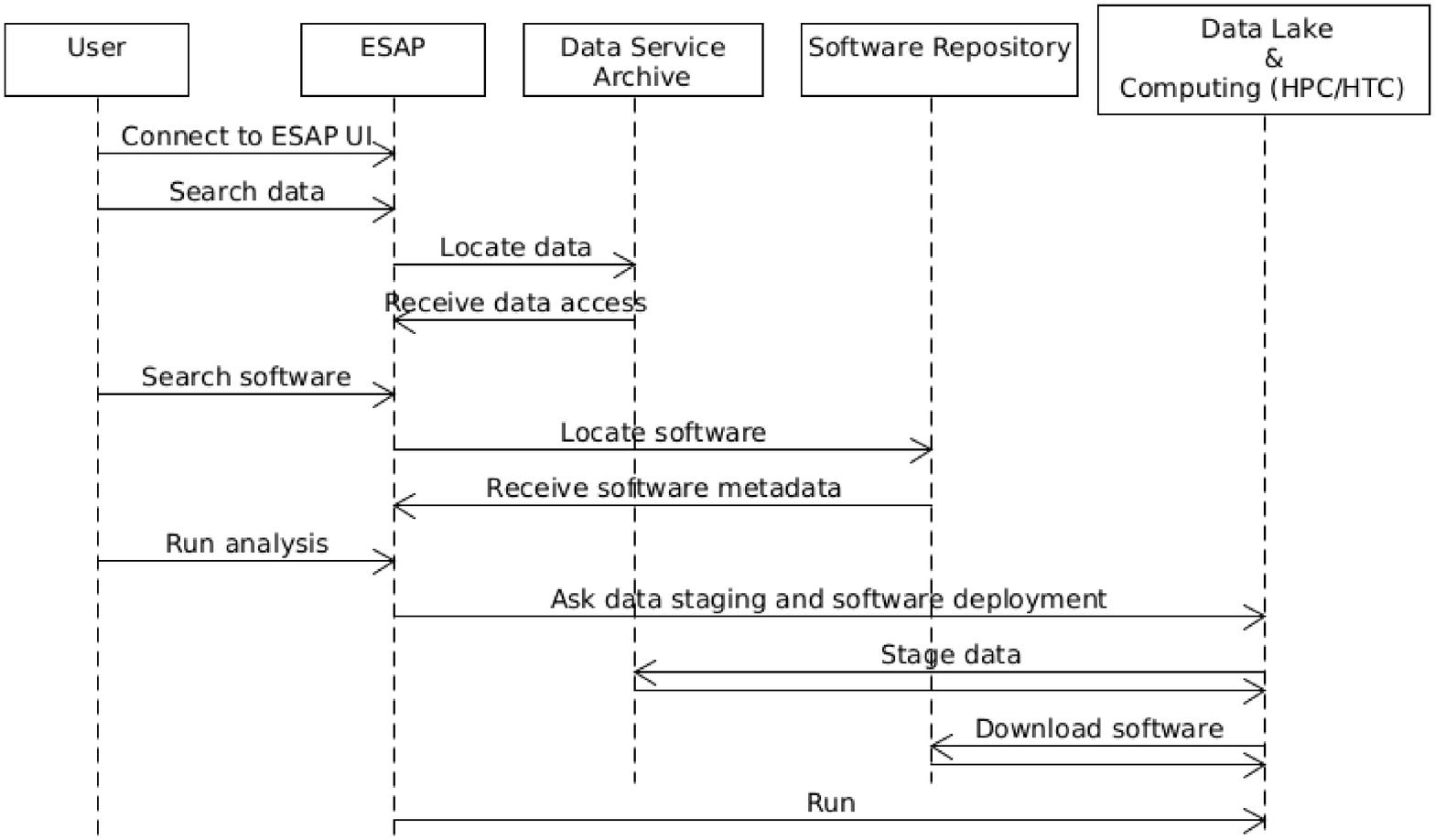}{use_case_fig1}{{\bf First scenario:} the data analysis performs in computing resources associated to the data lake (see sec.~\ref{datamng}), in the same infrastructure.}
\articlefigure{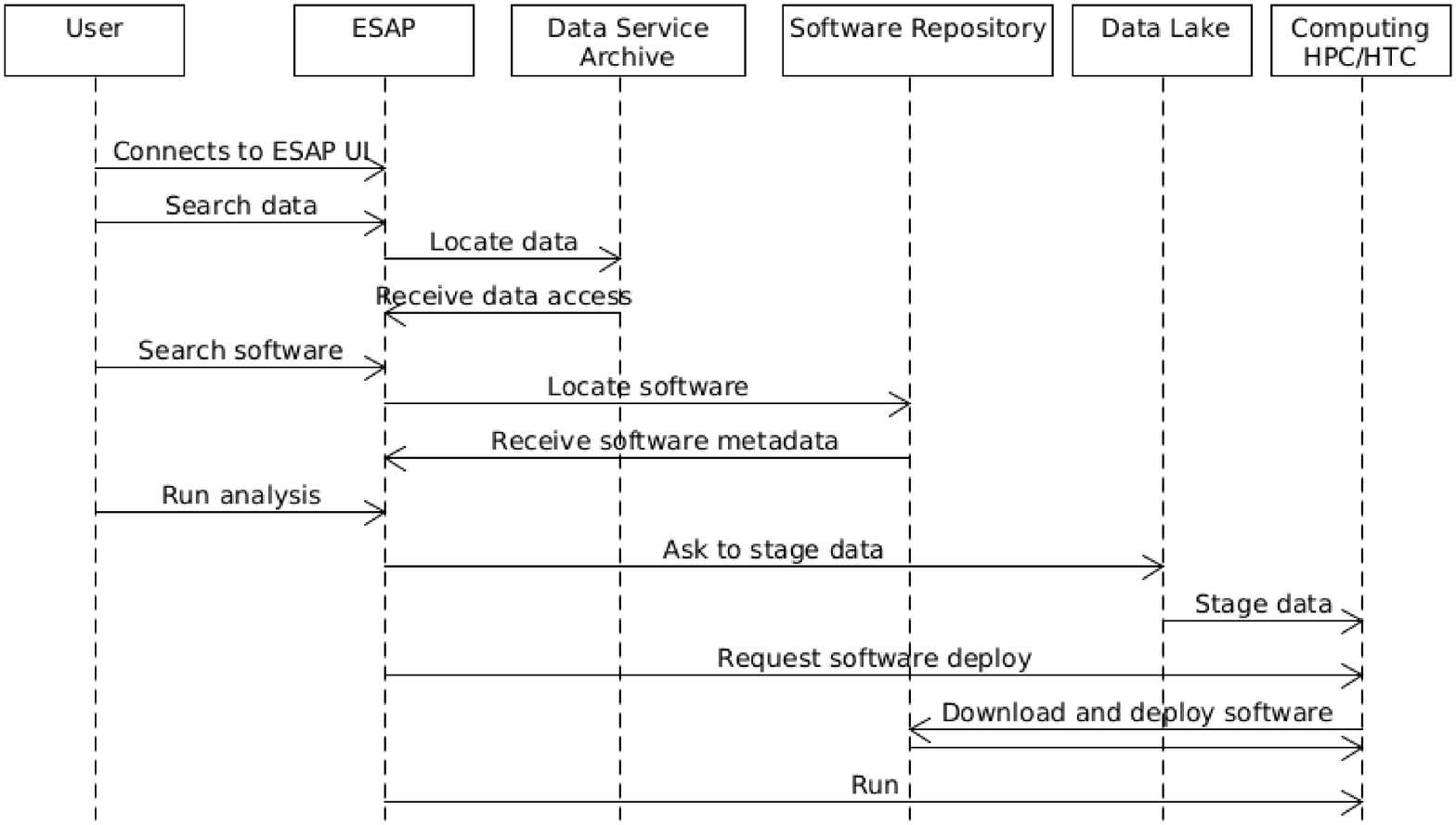}{use_case_fig2}{{\bf Second scenario:} the data analysis performs in computing resources not associated to the data lake, i.e. in a computing center different from which hosting the data lake (see sec.~\ref{datamng}).}
\section{HPC and HTC Integration}
For many ESFRIs and Research Infrastructures, the data scales involved require significant computational resources (storage and compute) to support processing and analysis. The EOSC-ESFRI science platform therefore must implement appropriate interfaces to an underlying HPC (High Performance Computing) or HTC (High Throughput Computing) infrastructure to take advantage of it. ESAP aims to provide users with the ability to access data and to deploy user-initiated processing and analysis tasks on  HTC and HPC infrastructures both in batch mode and maintaining interactivity. Responsiveness in the analysis system will be a challenge.
It is crucial to analyse and to identify the main requirements in terms of authentication and authorization policies, data management, computation resources management and software run to successfully integrate in ESAP HPC and HTC computation resources access tools.
\section{Authentication and Authorization}
Astronomical data are generally public, sometimes after a relatively short embargo period. For this reason Authentication and Authorization to data access is often taken in to account as a secondary problem. The case of computing resources, in terms of both hardware and software, is different: computation infrastructures generally have usage policy, sometimes posed as base for billing rules, and analysis software can be licensed or protected by owner's exclusive rights. For this reason the integration of HTC and HPC resources in the ESAP requires to implement mechanisms for user's authentication (to know who is the user requesting to access resources) and authorization (to assign the accessible/purchased resources both storage and computation). Authentication and authorization issues must be taken into account and to be faced at two levels (see~\citet{AARC_bluprint}): ESFRI Science Analysis Platform level and Data and Computing Center level. The implementations at the two levels must be compliant.
\section{Data Management}
\label{datamng}
To analyze data in an computation infrastructure (HPC or HTC), both large dimension, like CINECA (see~\citet{CINECA}) and EGI (see~\citet{2018arXiv180711318O}) or small dimension, like INAF computation infrastructure (see ~\citet{IT-OATS, CHIPP}), they must be staged in a storage area associated to and accessible by the infrastructure. 
Through ESAP the user can search and select data (See~\citet{P1-59_adassxxx}).
In the environment of the ESCAPE project, a prototype of an open access data storage for the ESCAPE community has been set up: the ESCAPE Data Infrastructure for Open Science (DIOS), also called data lake. Data downloaded from a public web server can be transferred into the data lake and their path is exposed to users, using the RUCIO data management service. 
The integration of HPC and HTC infrastructures in ESAP requires to exploit these features: data selected through ESAP must be staged in the DIOS and, when needed for analysis, transferred in a target computing center exploiting RUCIO capabilities. 
\section{Workflow management}
The user who wants to analyze data exploiting an HPC or HTC infrastructure has the need to
deploy analysis software in the target infrastructure where data must be available. To satisfy the different scenarios depicted above a task description is needed. It should
contain: the user identity and rights of access to the data and to the software; the data location and, if needed, the transfer information (protocols and tools); the analysis software identifier, the location and, if needed, the transfer information. The environment must be
suitably configured in case of not self-contained software. Containers, for example, could be used to simplify this task. 
\section{Conclusions} 
The analysis work carried out touches all the main aspects of the integration, but must be deepened from the point of view of the interaction with other project features: Data Lake and Software Repository.

\acknowledgments This work benefits support from the projects ESCAPE (grant n.~824064) and EuroEXA FET-HPC (grant n.~754337), funded by the Eu\-ro\-pean Commission, Horizon2020 programme.
\bibliography{P1-255}  


\end{document}